\documentclass[aps,prb,twocolumn,superscriptaddress,longbibliography]{revtex4-1}
\usepackage{graphicx}
\usepackage[colorlinks=true,citecolor=blue]{hyperref}
\usepackage{amsmath}
\usepackage{amssymb}
\usepackage{soul}

\begin{document}

\title{Artificial heavy fermions in a van der Waals heterostructure}

\author{Viliam Va\v{n}o}
\author{Mohammad Amini}
\author{Somesh C. Ganguli}
\author{Guangze Chen}
\author{Jose L. Lado}
\email{jose.lado@aalto.fi}
\author{Shawulienu Kezilebieke}
\author{Peter Liljeroth}
\email{peter.liljeroth@aalto.fi}

\affiliation{Department of Applied Physics, Aalto University, FI-00076 Aalto, Finland}

\date{\today}

\maketitle

\textbf{Heavy fermion systems represent one of the paradigmatic strongly correlated states of matter \cite{RevModPhys.56.755,Annurev_2016_Yazdani,Annurev_2016_Coleman,Wirth2016,Jiao2020}. They have been used as a platform for investigating exotic behavior ranging from quantum criticality and non-Fermi liquid behavior to unconventional topological superconductivity \cite{Pfleiderer2004,RevModPhys.73.797,RevModPhys.79.1015,Gegenwart2008,Aynajian2012,Allan2013,Zhou2013,Wirth2016,Jiao2020}. Heavy fermions arise from the exchange interaction between localized magnetic moments and conduction electrons that leads to the well-known Kondo effect \cite{PhysRev.124.41,Kondo1964,Kouwenhoven_2001}. In a Kondo lattice, the interaction between the localized moments gives rise to a band with heavy effective mass. This intriguing phenomenology has so far only been realized in compounds containing rare-earth elements with 4f or 5f electrons \cite{RevModPhys.56.755,Wirth2016,PhysRevB.103.085128,ramires2021emulating}. Here, we realize a designer van der Waals heterostructure where artificial heavy fermions emerge from the Kondo coupling between a lattice of localized magnetic moments and itinerant electrons in a 1T/1H-TaS$_2$ heterostructure. We study the heterostructure using scanning tunneling microscopy (STM) and spectroscopy (STS) and show that depending on the stacking order of the monolayers, we can either reveal the localized magnetic moments and the associated Kondo effect, or the conduction electrons with a heavy-fermion hybridization gap. Our experiments realize an ultimately tuneable platform 
for future experiments probing enhanced many-body correlations, dimensional tuning of quantum criticality, and unconventional superconductivity in two-dimensional artificial heavy-fermion systems \cite{Neumann1356,Shishido980,Mizukami2011}.}

\begin{figure}[t!]
    \centering
    \includegraphics[width=\columnwidth]{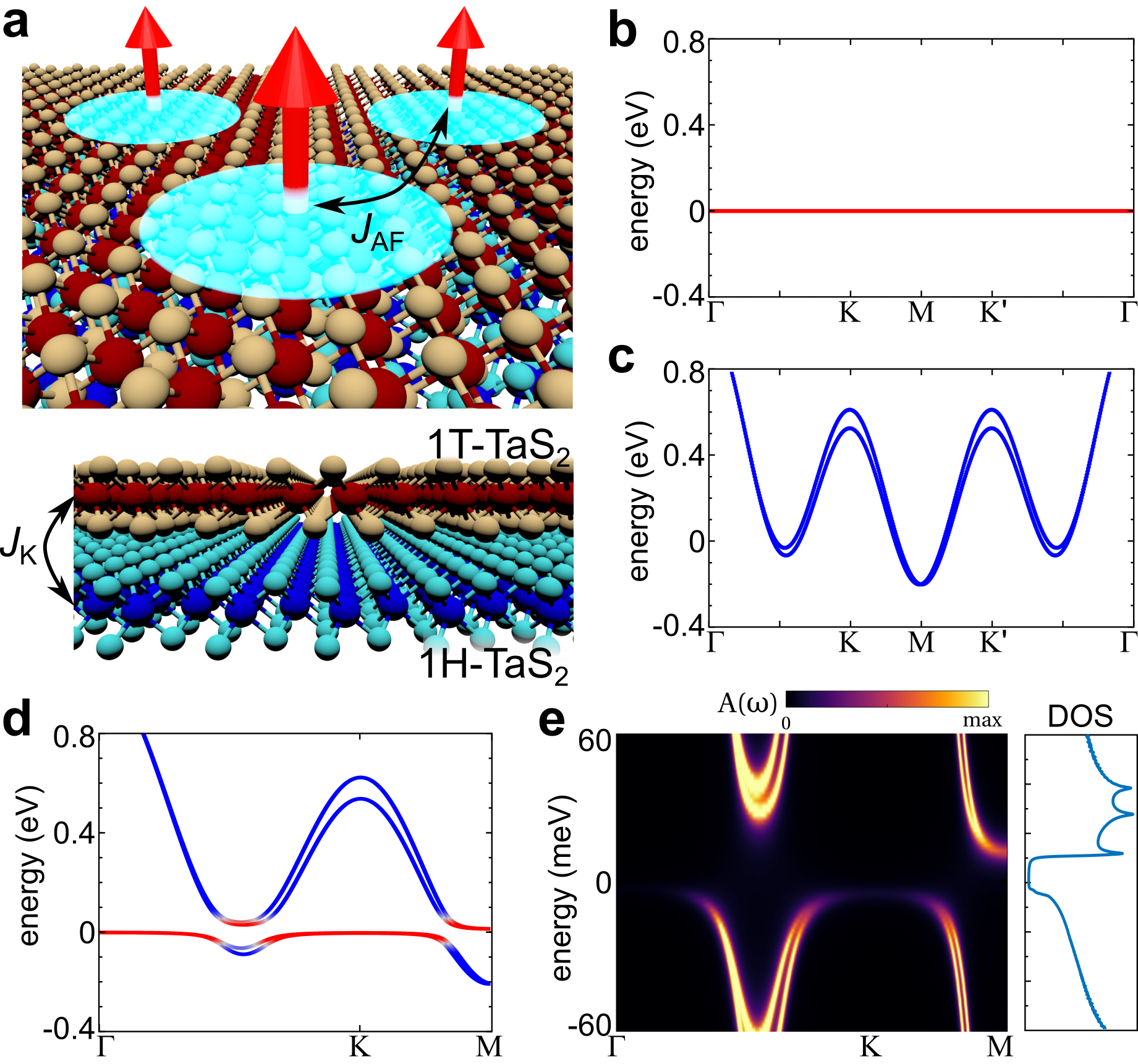}
	\caption{\textbf{Artificial heavy fermion heterostructure.} \textbf{a}, Sketch of the heterostructure and scheme of the localized moments in 1T- and itinerant electrons in 1H-layers, respectively. The layers are coupled through exchange coupling $J_K$. \textbf{b,c}, Sketches of the pseudo-fermion (b) and electron (c) band structures of the 1T- and 1H-layers. \textbf{d}, Band structure in the heavy fermion regime showing the emergence of a gap. The colors refer to the projected density of states to 1T pseudo-fermion (red) and 1H electrons (blue). \textbf{e}, Calculated spectral function and corresponding density of states (DOS) of the 1H-layer in the 1H/1T-bilayer heterostructure close to the Fermi level.}
    \label{fig:schem}
\end{figure}

\begin{figure*}[t!]
    \centering
    \includegraphics[width=.8\textwidth]{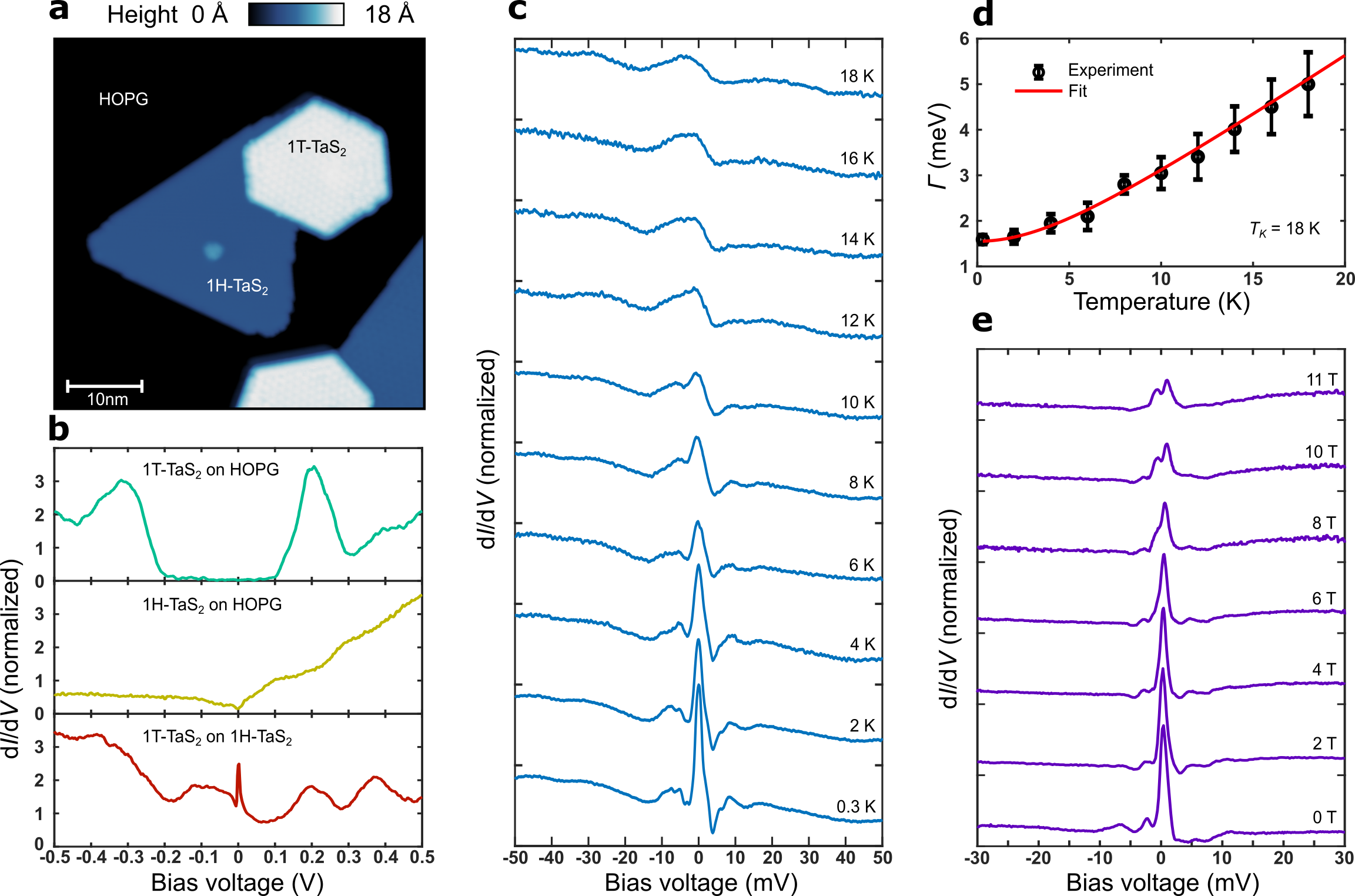}
	\caption{\textbf{Kondo resonance in a 1T/1H-TaS$\mathbf{_2}$ vertical heterostructure.} \textbf{a}, STM image of a 1T/1H-TaS$_2$ vertical heterostructure on HOPG substrate, taken at $V = 1$ V and $I$ = $20$ pA. \textbf{b}, Tunneling spectra of ML 1T-TaS$_2$ on HOPG (green), 1H-TaS$_2$ on HOPG (yellow) and ML 1T-TaS$_2$ on 1H-TaS$_2$ (red). \textbf{c}, Tunneling spectroscopy of the Kondo resonance measured on a 1T/1H-TaS$_2$ vertical heterostructure at different temperatures, the spectra are vertically offset for clarity. \textbf{d}, Temperature dependence of the intrinsic Kondo resonance width (black circles with error bars) obtained by fitting the spectra in (c). Fit (red line) yields a Kondo temperature of 18K. \textbf{e}, Magnetic field dependence of the Kondo resonance in tunneling spectroscopy.}
    \label{fig:Kondo}
\end{figure*}
In recent years, van der Waals (vdW) heterostructures have become the de-facto platform to engineer artificial electronic phenomena \cite{Geim2013,Liu2016,Novoselovaac9439,Kennes2021,Andrei2021}. They have only weak interactions between the different layers, which allows each layer to retain its intrinsic properties. The interfaces between the layers are essentially contamination and defect-free and well-defined down to the atomic level. These factors make it possible to combine materials with seemingly competing electronic orders. For the realization of artificial heavy fermions, we need to couple a material hosting local moments with an itinerant electron bath \cite{ramires2021emulating,PhysRevB.103.085128}. In particular, some monolayer transition metal dichalcogenides (TMDs) 
show metallic behaviour \cite{Law6996,Ugeda2015,delaBarrera2018,Zhao2019}, while others 
are known to realize a correlated, charge-density wave driven state hosting local magnetic moments and potentially a quantum spin liquid state \cite{Law6996,Cho2016,TaS2_Mottness,Chen2020,ruan2020imaging}. 

Here, we use molecular beam epitaxy (MBE) to grow bilayer 1T-TaS$_2$ / 1H-TaS$_2$ heterostructures (Fig.~\ref{fig:schem}a) and characterize them using low-temperature scanning tunneling microscopy (STM) and spectroscopy (STS). We demonstrate that the localized moments in 1T-TaS$_2$ Kondo couple with the itinerant electrons of the Ta d-band in 1H-TaS$_2$ and give rise to a heavy-fermion band structure. The use of vdW materials allows unprecedented external control using e.g.~light and electrostatic gating that cannot be reached in the traditional heavy-fermion systems. These results mark the latest advance in realizing strongly correlated states in vdW heterostructures and open up a pathway towards even more exotic phases such as artificial heavy-fermion superconductivity in the future.

1T-TaS$_2$ exhibits a charge-density-wave (CDW) state that results in a large unit cell hosting a single localized moment \cite{Law6996,Cho2016,TaS2_Mottness}. This local moment can be exchange-coupled with the Ta d-band of the underlying 1H-TaS$_2$ layer forming a Kondo ground state. When these Kondo impurities form a lattice (defined by the CDW unit cell), it results in a pseudo-fermion flat band in the bilayer electronic structure. This flat band (Fig.~\ref{fig:schem}b) has an avoided crossing with the electronic dispersive band of the 1H-TaS$_2$ layer (Fig.~\ref{fig:schem}c) due to the Kondo coupling, in an exact analog of a natural heavy-fermion compound (Fig.~\ref{fig:schem}d). As a consequence of the previous picture, we would expect to observe the Kondo resonances when probing the local density of states (LDOS) of the 1T-TaS$_2$ layer. On the other hand, the LDOS on the 1H-TaS$_2$ side should exhibit a signature of a heavy-fermion hybridization gap at the Fermi level (Fig.~\ref{fig:schem}e). We will show in the following that we can realize precisely this response in our experiments.

We have tuned our MBE growth to produce both 1T and 1H monolayers as well as bilayer islands with in all possible combinations: 1H and 1T bilayers, and 1H/1T and 1T/1H heterobilayers (details of the sample growth and STM experiments are given in the Methods section). Fig.~\ref{fig:Kondo} shows results of STM and STS experiments of a 1T/1H-TaS$_2$ vertical heterostructure. The tunneling spectroscopy matches with the expected electronic structure of the 1T- and 1H-TaS$_2$ (Fig.~\ref{fig:Kondo}b) \cite{Law6996,Cho2016,TaS2_Mottness,delaBarrera2018}. 1T-TaS$_2$ (green curve) is a Mott insulator with upper Hubbard band at $V = 0.2$ V, lower Hubbard band at $V = 0.3$ V and a charge gap of $\sim0.3$ V, while 1H-TaS$_2$ (yellow curve) is a metal with finite LDOS at the Fermi level. 
The LDOS of the 1T-TaS$_2$ on top of 1H-TaS$_2$ (red curve) has a strong peak at the Fermi level, which is consistent with the Kondo lattice effect and with the previous experiments on an analogous heterostructure \cite{ruan2020imaging}.

\begin{figure*}[t!]
    \centering
    \includegraphics[width=.8\textwidth]{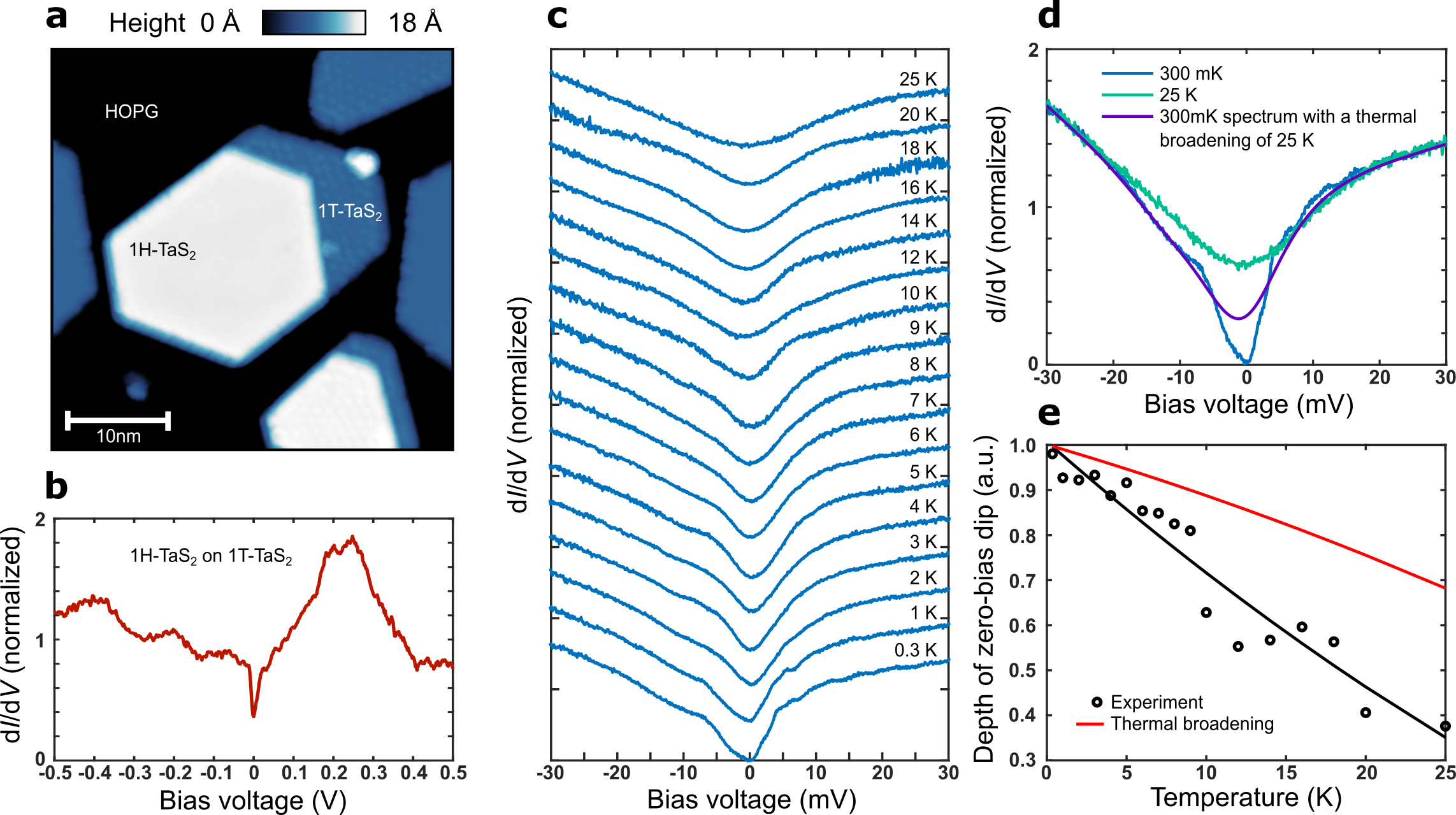}
	\caption{\textbf{Heavy fermion hybridization gap in a 1H/1T-TaS$\mathbf{_2}$ vertical heterostructure.} \textbf{a}, STM image of a 1H/1T-TaS$_2$ vertical heterostructure on HOPG substrate, obtained at $V = 1$ V and $I$ = $8$ pA. \textbf{b}, Tunneling spectrum of 1H-TaS$_2$ on a ML 1T-TaS$_2$. \textbf{c}, Tunneling spectroscopy of a heavy fermion gap measured on a 1H/1T-TaS$_2$ vertical heterostructure at different temperatures, the spectra are vertically offset for clarity. \textbf{d}, Tunneling spectra of the HF gap measured at 300mK (blue) and 25K (green). The purple curve corresponds to a tunneling spectrum measured at 300mK with a thermal broadening of 25K. \textbf{e}, Temperature evolution of the zero-bias dip depth (black circles) with an eye-guide (black line). Red line is a simulated temperature evolution acquired by applying thermal broadening to the 300mK spectrum.}
    \label{fig:HFgap}
\end{figure*}

In order to confirm the Kondo nature of the peak in the tunneling spectroscopy at zero bias, we have measured its evolution with temperature and applied magnetic field \cite{Zhang2013}. The peak at zero bias is getting gradually wider and lower in intensity with temperature (Fig.~\ref{fig:Kondo}c), which cannot be accounted for by the effect of thermal broadening. The tunneling spectra at different temperatures were also fit to a Fano lineshape (see Supplementary Information), resulting in a temperature-dependence of the Kondo resonance width $\Gamma$ (Fig.~\ref{fig:Kondo}d, black circles with error bars). This width follows a well-known temperature dependence and a fit (red line) yields a Kondo temperature of $T_K = 18$ K. Applying a magnetic field perpendicular to the sample surface broadens the zero-bias peak at lower magnetic fields and eventually splits it at high magnetic fields (Fig.~\ref{fig:Kondo}e), a behavior expected for the Zeeman splitting of a Kondo resonance\cite{Zhang2013}. 

Scanning tunneling spectroscopy of this heterostructure probes the LDOS of the top layer. As shown above, the 1T-TaS$_2$ on top of 1H-TaS$_2$ develops a Kondo resonance at the Fermi level. When the layers in the heterostructure are inverted, i.e. with 1H-TaS$_2$ on top of 1T-TaS$_2$, we observe a $\sim7$ mV gap around the Fermi level (see Fig.~\ref{fig:HFgap}). Fig.~\ref{fig:HFgap}b shows a large bias tunneling spectroscopy of the 1H/1T-TaS$_2$ with a dip at zero bias. The tunneling conductance at zero bias is finite because of larger lock-in bias modulation used for large-bias spectroscopy. However, by performing higher-resolution tunneling spectroscopy, we find out the tunneling conductance at zero bias is zero at the lowest temperature (see Fig.~\ref{fig:HFgap}c). This is not the case for 1H-TaS$_2$ on HOPG or for bilayer  1H-TaS$_2$, even though both of these develop a dip around the Fermi level (see Supplementary Information). This dip is a common feature of ML metallic TMDs \cite{Ugeda2015, Ryu2018}.

As can be seen in Fig.~\ref{fig:HFgap}c, the gap continuously closes with increasing temperature, which is consistent with previous experiments measuring the temperature-dependence of a heavy-fermion hybridization gap \cite{Ernst2011, Rler2014, Aynajian2012}. Once again, this gap closing cannot be accounted for only by the thermal broadening of the tunneling spectroscopy. This can be seen in Fig.~\ref{fig:HFgap}d, where we take the lowest temperature tunneling spectrum (blue curve) and simulate the same spectrum with a thermal broadening corresponding to $T=25$ K (purple curve). This thermally broadened spectrum is significantly different from the measured one at 25 K (green curve), which is due to the combined effect of thermal broadening and the closing of the $\sim7$ mV gap. The difference between simulated thermal broadening of the lowest temperature spectrum and the actually measured temperature dependence can also be seen in the temperature dependence of a zero-bias tunneling conductance. This is shown in Fig.~\ref{fig:HFgap}e, where we look at the temperature dependence of the depth of the zero-bias dip. In the experimental data, the features wash out with increasing temperature much more rapidly than in the case of simulated thermal broadening.

Besides the heavy-fermion hybridization gap, there are also other possible explanations for the gap in tunneling spectroscopy of a 1H/1T-TaS$_2$ heterostructure that we need to rule out. The first possibility is a Coulomb gap, where the repulsive Coulomb interactions of quasiparticles confined in an island would lead to a gap around the Fermi level \cite{vonDelft2001,Bose2014,ganguli2020controlling,Kouwenhoven2001,Ihn2009}. The repulsive Coulomb interactions should increase with decreasing island size; however, we do not observe a correlation between the gap width and island size in our experiments (see Supplementary Information). Another possibility would be a CDW gap. Even though the 1H-TaS$_2$ exhibits a $3\times3$ charge density wave, in the case of 1H-TaS$_2$ on top of 1T-TaS$_2$, we observe no CDW modulation in STM images with atomic resolution (see Supplementary Information). 
Finally, the observed gap could also in principle be of a superconducting origin. 1H-TaS$_2$ is already a superconductor \cite{delaBarrera2018}, and by creating a heterostructure one could increase the electron-phonon interaction, or an unconventional magnon-mediated superconductivity could arise in this heterostructure\cite{She2017}. Such a two-dimensional superconductivity should be strongly influenced by applying out of plane magnetic field, and we see no significant changes to the observed gap up to 10 T (see Supplementary Information). Overall, we can rule out all these options and conclude that the observed gap is a heavy-fermion hybridization gap.

In conclusion, we have realized an artificial van der Waals heterostructure hosting heavy fermion physics. Our STM and STS experiments demonstrate the presence of the fundamental ingredients for this, Kondo screening of local moments and development of a heavy-fermion gap that emerges in our heterostructures. These results open a pathway towards a whole new family of two-dimensional materials showing the emergence of a strongly correlated state of matter that previously required rare earth elements. The use of van der Waals materials allows for an unprecedented level of control over the system parameters that is not available in rare-earth compounds, such as changing the twist angle or tuning the chemical potential by the gating of a sample. Ultimately, this will enable the study of heavy fermion superconductivity, quantum criticality and non-Fermi liquid phases tuneable by gating and twist engineering.

\section*{Methods}
\textbf{Sample preparation.} 
TaS$_2$ was grown by molecular beam epitaxy (MBE) on highly oriented pyrolytic graphite (HOPG) under ultra-high vacuum conditions (UHV, base pressure $\sim1\times10^{-10}$ mbar). HOPG crystal was cleaved and subsequently out-gassed at $\sim800^\circ$C. High-purity Ta and S were evaporated from an electron-beam evaporator and a Knudsen cell \cite{Hall2018}, respectively. Prior to growth, the flux of Ta was calibrated on a Au(111) at $\sim1$ monolayer per hour. The ratio of 1T to 1H-TaS$_2$ can be controlled via substrate temperature and the overall coverage \cite{Lin2018}. Before the growth, HOPG substrate temperature was stabilised at $\sim680^\circ$C. The sample was grown in a S pressure of $\sim5\times10^{-8}$ mbar and the growth duration was 25 minutes.

\textbf{STM measurements.} After the sample preparation, it was inserted into the low-temperature STM (Unisoku USM-1300) housed in the same UHV system and all subsequent experiments were performed at $T = 300$ mK. STM images were taken in the constant-current mode. d$I$/d$V$ spectra were recorded by standard lock-in detection while sweeping the sample bias in an open feedback loop configuration, with a peak-to-peak bias modulation of 0.4-1 mV at a frequency of 731 Hz, a modulation of 5 mV was used for d$I$/d$V$ spectra at larger biases.

\section*{Acknowledgements}
This research made use of the Aalto Nanomicroscopy Center (Aalto NMC) facilities and was supported by the European Research Council (ERC-2017-AdG no.~788185 ``Artificial Designer Materials'') and Academy of Finland (Academy professor funding nos.~318995 and 320555, Academy postdoctoral researcher no.~309975, Academy research fellow nos.~331342 and 336243). We acknowledge the computational resources provided by the Aalto Science-IT project. 

\bibliography{biblio}

\end{document}